# Entanglement sudden birth and sudden death in a system of two distant atoms coupled via an optical element

Maryam Ashrafi & M.H. Naderi

# Entanglement sudden birth and sudden death in a system of two distant atoms coupled via an optical element

Maryam Ashrafi[a] and M.H. Naderi[a,b]*

[a]*Department of Physics, Faculty of Science, University of Isfahan, Hezar Jerib, 81746-7344, Isfahan, Iran;* [b]*Quantum Optics Group, Department of Physics, Faculty of Science, University of Isfahan, Hezar Jerib, 81746-7344, Isfahan, Iran*

An investigation is reported of the collective effects and the dynamics of atom–atom entanglement in a system of two distant two-level atoms which are coupled via an optical element. In the system under consideration, the two atoms, which are trapped in the foci of a lens, are coupled to a common environment being in the vacuum state and they emit photons spontaneously. A fraction of the emitted photons from each atom is thus focused on the position of the other atom. The presence of optical element between two distant atoms leads to the occurrence of delayed collective effects, such as delayed dipole–dipole interaction and delayed collective spontaneous emission, which play the crucial role in the dynamical behaviour of the entanglement. We discuss the phenomena of entanglement sudden birth, entanglement sudden death, and revival of entanglement for both cases of initial one-photon and initial two-photon unentangled atomic states. We show that the evolution of the entanglement is sensitive not only to the interatomic distance but also to the initial state of the system as well as to the properties of the optical element.

## 1. Introduction

Quantum entanglement, as one of the most profound features of quantum mechanics, plays a central role in quantum information science and quantum computation [1,2]. It is an essential resource for implementation of many quantum information protocols, such as quantum computation [3], quantum teleportation [4], quantum key distribution [5,6], controlled quantum logic [7], quantum cryptography [5,8], and superdense coding [9]. Entanglement generation has been extensively investigated in a variety of physical systems, ranging from quantum optical to condensed matter systems [10–14].

Of particular interest is the generation of entangled states in two-atom systems, since they can represent two qubits, the building blocks of the quantum gates that are essential to implement quantum protocols in quantum information processing. It has been shown that entangled states in a two-atom system can be generated by continuous driving of the atoms with a coherent or chaotic thermal field [15–17], or by a pulse excitation followed by a continuous observation of radiative decay [18–20]. However, the coupling of the quantum system to the environment usually leads to a dissipative evolution of quantum coherence (decoherence) and loss of the useful entanglement. Thus, the study of dynamical evolution of two entangled qubits coupled to environmental degrees of freedom is of fundamental importance in quantum information sciences. A typical source of decoherence is spontaneous emission resulting from the interaction of a system with an external environment. In composed systems there are two kinds of coupling between the system and its environment. In the first type each part of the system is coupled with own environment while in the second type all parts of the system coupled with a common environment. In the second type, collective effects in the system may lead to the creation of entanglement between two disentangled qubits [21,22]. This effect is called the environment-induced entanglement [17,23].

The time evolution of entanglement for a system of two qubits or two two-level atoms can be described qualitatively for various physical situations, and it has been studied extensively in recent years [24–34]. The destructive effect of spontaneous emission on entanglement encoded into qubits can take different time scales. The decoherence time depends on the damping rate of the state in which the entanglement was initially encoded and usually the decay process induced by spontaneous emission occurs exponentially in time. A lot of discussion has been devoted to the problem of disentanglement of the two-qubit system in a finite time that is much shorter than the exponential decoherence time of spontaneous emission. This interesting phenomenon has been termed entanglement sudden death (ESD) [27–30]. It appears on bipartite subsystems each interacting with its own environment, depending on the initial bipartite state. ESD has recently been observed in two elegantly designed experiments with photonic qubits [31,32] and atomic ensembles [33].

---

*Corresponding author. Email: mhnaderi@phys.ui.ac.ir

Although the sudden death feature is concerned with the disentangled properties of spontaneous emission there is an interesting 'sudden' feature in the temporal creation of entanglement from initially independent qubits [34,35]. The phenomenon is termed entanglement sudden birth (ESB), as it is opposite to the sudden death of entanglement and arises dynamically during the spontaneous evolution of an initially separate qubits. ESB has been studied for identical qubits coupled to either a common multimode vacuum field [23,34,36] or to a damped single-mode cavity field [37,38]. ESD and ESB have also been discussed for a two-atom system interacting with a common structured reservoir [39], for a system of two two-level atoms interacting with a common Markovian reservoir at finite temperature [40], and for a system of two initially entangled qubits interacting independently with two uncorrelated reservoirs at zero temperature beyond the Markovian approximation [41]. However, most abovementioned studies about ESB require two close-lying atoms with a distance $d$ of the order of the wavelength $\lambda$ of light emitted by the atom or smaller such that they can undergo a correlated decay resulting from the collective properties of the system. The collective effects depend strongly on the distance between the two atoms and rapidly vanish with increasing the interatomic distance [42]. Recently, it has been proposed to realize ESB between two distant qubits (about 10 wavelengths apart) by using left-handed materials [43] via enhancement of the interaction between distant qubits. The entanglement of two qubits mediated by metal nanowire [44] and a one-dimensional plasmonic waveguide [45] has also been studied. In [46], the generation of entanglement between two two-level atoms, separated by a slab of materials (including metal and metamaterials), via surface modes has been discussed. Moreover, studies have also been focused on achieving strong coupling between two distant atoms by means of optical elements, such as lenses of large numerical aperture [47–49] or optical fibers [50]. When the photonic interaction between the atoms is mediated by an optical element, its strength is characterized by the fraction $\kappa$ of modes of the electromagnetic field which propagate from one atom to the other via the optical element. Thus $\kappa$ replaces the scaling with $\lambda/d$ of the free-space case, and coupling over much larger distances than $\lambda$ may be achieved.

The purpose of the present paper is to investigate the dynamics of entanglement, measured by concurrence, for a system of two distant atoms $(d > \lambda)$ that are coupled by radiation via an optical element. The optical element collects a fraction of the radiation emitted by each atom and focuses it onto the other one. We assume that both atoms interact with a common environment being in the vacuum state. In [42], it has been shown that in a system of two close atoms without optical element, when the atoms are coupled with a common external environment, the collective interactions between the atoms give rise not only to modified dissipative spontaneous emission but also to a coherent (dipole–dipole) interaction between the atoms. With increasing interatomic distance, such that it becomes larger than the transition wavelength of the atoms, one can neglect the free-space dipole–dipole interaction between the atoms. However, we find that in a system of two distant atoms, the presence of the optical element leads to a coherent coupling between the two atoms. This coupling is analogous to the dipole-dipole interaction in a system of two close atoms without the optical element. This paper is organized as follows. In Section 2 we introduce the physical model applied in the paper. In Section 3 we study the dynamical behavior of the atom–atom entanglement for two special cases of the initial conditions. In the first, one of the atoms is in its excited state and the other is assumed to reside in its ground state. In the second, both atoms are assumed to reside in their excited states. Finally, in Section 4, we summarize our conclusions.

## 2. Physical model

In this section, we describe the theoretical model for the system under consideration. For this purpose, we use the theoretical formalism presented in [51] for one atom in front of a mirror, and generalize it to the case of two coupled atoms.

### 2.1. System Hamiltonian

We consider a system of two identical two-level atoms with the excited state $|e\rangle$ and the ground state $|g\rangle$ connected by a dipole transition with dipole moment $\vec{p}$ and transition frequency $\omega_0$. The atoms are located at fixed positions $\vec{r}_1$ and $\vec{r}_2$. We consider the case in which the interatomic distance $d = |\vec{r}_1 - \vec{r}_2|$ is larger than the wavelength $\lambda = 2\pi c/\omega_0$; hence the free-space dipole–dipole interaction between the atoms can be neglected. As sketched in Figure 1, we assume that a lens with the focal length $F$ is placed between the atoms such that they are located at the focal points. In [47], a similar situation was realized, coupling two atoms via a mirror and a lens. The lens focuses a

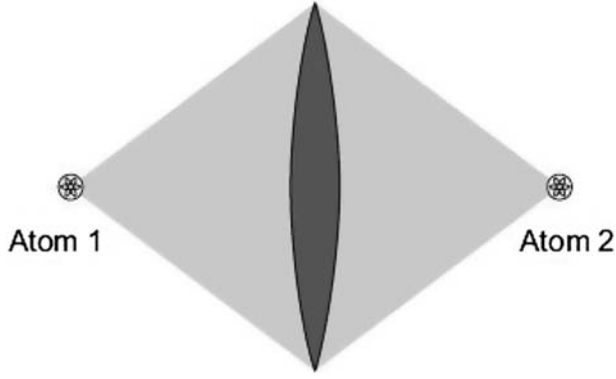

Figure 1. Schematic of a system of two distant atoms coupled by a lens focusing part of their emitted radiation onto each other.

fraction of the radiation emitted by one atom onto the other. We label these modes (internal modes) with $v$, in order to distinguish them from the external modes, labeled with $\mu$, which do not couple the atoms. The strength of the photonic coupling between the atoms mediated by the lens is defined through the fraction of $4\pi$ solid angle within which the radiation from one atom is focused onto the other. Denoting this fraction with $\delta\Omega_0$, the strength of the coupling is given by the dimensionless parameter $\kappa$ [49]

$$\kappa = \frac{3}{8\pi}\int_{\delta\Omega_0} d\Omega(1 - |\vec{p}.\vec{k}_v|^2/|\vec{p}|^2|\vec{k}_v|^2) \quad 0 < \kappa < 1, \quad (1)$$

where $\vec{k}_v$ is the wavevector of the internal mode $v$. The limit $\kappa \to 0$ corresponds to the case without the lens and $\kappa \to 1$ would describe an ideal lens that maps all radiation from one atom onto the other.

The Hamiltonian of the system is given by

$$H = H_0 + H_I, \quad (2)$$

where

$$H_0 = \sum_{i=1,2} \hbar\omega_0 \sigma_i^+ \sigma_i^- + \sum_{q=\mu,v} \hbar\omega_q a_q^\dagger a_q, \quad (3)$$

denotes the free Hamiltonian of the system and

$$H_I = -i\hbar \sum_{i=1,2}\sum_{q=\mu,v} g_q(\vec{r}_i)e^{i\vec{k}_q.\vec{r}_i}\sigma_i^+ a_q + H.c., \quad (4)$$

is the interaction Hamiltonian under the rotating wave approximation. Here, the radiation mode $q$ of frequency $\omega_q$, wave vector $\vec{k}_q$ and polarization $\hat{e}_q \perp \vec{k}_q$ is described by the bosonic annihilation and creation operators $a_q$ and $a_q^\dagger$, $\sigma_i^+ (\sigma_i^-)$ denotes the atomic raising (lowering) operator of the $i$th atom, and the coupling constant $g_q(\vec{r}_i)$ is given by

$$g_q = \sqrt{\frac{\omega_q}{2\hbar\varepsilon_0 V_q}}(\vec{p}.\hat{e}_q), \quad (5)$$

with the quantization volume

$$V_q = \frac{1}{\int_{\delta\Omega_q} d^3k}. \quad (6)$$

For the internal modes $\delta\Omega_q = \delta\Omega_0$ and for the external modes $\delta\Omega_q = 4\pi - \delta\Omega_0$.

### 2.2. Master equation

In order to describe the dynamics of the system under consideration we consider the equation of motion for the density operator of the combined system $\rho_{AF}^I$ in the interaction picture

$$\frac{\partial \rho_{AF}^I}{\partial t} = \frac{1}{i\hbar}[V_I(t), \rho_{AF}^I(t)], \quad (7)$$

where

$$V_I(t) = e^{iH_0 t/\hbar} H_I e^{-iH_0 t/\hbar}, \quad (8)$$

and

$$\rho_{AF}^I(t) = e^{iH_0 t/\hbar} \rho_{AF} e^{-iH_0 t/\hbar}. \quad (9)$$

We choose the initial state of the system such that there is no correlation between the atomic subsystem and the field modes, i.e.

$$\rho_{AF}^I(0) = \rho_A^I(0) \otimes \rho_F^I(0). \quad (10)$$

Furthermore, we suppose that the coupling between the atomic system and the field modes to be weak and there is no back reaction from the atomic system on the field [52]. Thus, we can write the density operator $\rho_{AF}^I(t)$ as

$$\rho_{AF}^I(t) = \rho_A^I(t) \otimes \rho_F^I(0). \quad (11)$$

By ignoring the interaction between the atoms through external modes and assuming the field modes being prepared in the vacuum state, the reduced master equation for the atomic subsystem is obtained as follows

$$\frac{\partial \rho}{\partial t} = \sum_{i=1,2} \sum_{\mu} g_{\mu}^2$$
$$\int_0^t dt' e^{i(\omega_0-\omega_\mu)(t-t')} [\sigma_i^+(t'), \sigma_i^-(t')\rho(t')]$$
$$+ \sum_{i,j=1,2} \sum_{\nu} g_{\nu}^2 e^{i\vec{k}_\nu \cdot (\vec{r}_i - \vec{r}_j)}$$
$$\int_0^t dt' e^{i(\omega_0-\omega_\nu)(t-t')} [\sigma_i^-(t'), \sigma_j^+(t')\rho(t')] + \text{H.c.}, \quad (12)$$

where we have used a shorter notation $\rho \equiv \rho_A^I$. According to the Fermat principle, in the presence of an optical element, all of the modes components take the minimum time for travelling from one atom to another one, so one can make the following replacement

$$e^{i\vec{k}\cdot(\vec{r}_i - \vec{r}_j)} \to e^{i\omega\tau}, \quad (13)$$

where $\tau = d/c$ is the propagation time for a photon from one atom to the other one via the optical element. By using this fact and changing the sums over $\mu$ and $\nu$ the reduced master Equation (12) takes the following form (see Appendix)

$$\frac{\partial \rho}{\partial t} = -\sum_{i,j=1,2} \frac{\Gamma_{ij}}{2} (\rho(t)\sigma_i^+(t)\sigma_j^-(t) + \sigma_i^+(t)\sigma_j^-(t)\rho(t)$$
$$- 2\sigma_i^-(t)\rho(t)\sigma_j^+(t)) - i\sum_{i \neq j}^2 \Omega_{ij}[\sigma_i^+(t)\sigma_j^-(t), \rho(t)], \quad (14)$$

where the diagonal term

$$\Gamma_{ii} \equiv \Gamma = \frac{|\vec{p}|^2 \omega_0^3}{3\hbar\pi\varepsilon_0 c^3}$$

is the free-space spontaneous emission rate, while the off-diagonal terms

$$\Gamma_{ij} = \kappa\Gamma\cos(\omega_0\tau)\Theta(t-\tau) \quad (i \neq j), \quad (15)$$

can be considered as the delayed collective spontaneous emission rates arising from the coupling between the atoms through the vacuum field in the presence of the optical element. Therefore, the spontaneous emission from one atom influences the spontaneous emission from the other. The second term on the right hand side of Equation (14) represents the delayed interaction between the atoms with the strength

$$\Omega_{ij} = \frac{\kappa\Gamma}{2}\sin(\omega_0\tau)\Theta(t-\tau) \quad (i \neq j). \quad (16)$$

Thus, the presence of optical element leads to a coherent coupling between the two atoms. This coupling is analogous to the dipole–dipole interaction in a system of two close atoms ($d \ll \lambda$) without the optical element [42]. In that system, the dipole–dipole interaction is induced by the vacuum field, while in the system under consideration the optical element is responsible for inducing the interaction. The collective parameters $\Gamma_{ij}$ and $\Omega_{ij}$ ($i \neq j$) change with the interatomic distance periodically and they are weighted by the parameter $\kappa$, showing that only a fraction of radiation emitted from each atom affect the other one. In a two-atom system without the optical element, the dipole–dipole interaction potential is given by $\Omega_{ij} \approx \frac{3\Gamma}{4(2\pi)^3}(\lambda/d)^3$ for $d \ll \lambda$, and $\Omega_{12} = 0$ for $d \gg \lambda$ [42]. As a numerical example, in the optical region we typically have $\omega_0 \approx 10^{15}$ Hz and $\Gamma \approx 10^7$ Hz. Then for $\Omega_{12} \approx \Gamma$, we obtain $d \approx 0.14\lambda$. However, Equations (15) and (16) show that the presence of the optical element leads to the revival of collective effects even when the atoms are separated by a distance $d$ much larger than the wavelength $\lambda$.

## 3. Entanglement in the system of two distant atoms

### 3.1. Master equation in the collective basis and its solution

The presence of coherent coupling between the two atoms suggests that the bare atomic states are no longer the eigenstates of the atomic system. In this case, the eigenstates and the corresponding eigenenergies of the system are as follows

$$|g\rangle = |g_1, g_2\rangle \qquad E_g = 0,$$
$$|s\rangle = \frac{1}{\sqrt{2}}(|e_1, g_2\rangle + |g_1, e_2\rangle) \qquad E_s = \hbar(\omega_0 + \Omega_{12}),$$
$$|a\rangle = \frac{1}{\sqrt{2}}(|e_1, g_2\rangle - |g_1, e_2\rangle) \qquad E_a = \hbar(\omega_0 - \Omega_{12}),$$
$$|e\rangle = |e_1, e_2\rangle \qquad E_e = 2\hbar\omega_0, \quad (17)$$

which are the collective Dick states [42]. The most important property of the collective states is that the symmetric state $|s\rangle$ and antisymmetric state $|a\rangle$ are maximally entangled states of the two-atom system. In the collective states representation, the two-atom system behaves as a single four-level system. As is seen, the energies of the symmetric and antisymmetric states depend on the interatomic distance and the properties of the optical element.

We now transform the master Equation (14) into the basis of the collective states of Equations (17). For this purpose, we use the collective operators $A_{ij} = |i\rangle\langle j|$, with $i, j = e, s, a, g$, that represent the energies ($i = j$) of the collective states and coherences ($i \neq j$). Thus, we obtain

$$\frac{\partial \rho}{\partial t} = \frac{1}{i\hbar}[H_s, \rho] + \left(\frac{\partial \rho}{\partial t}\right)_s + \left(\frac{\partial \rho}{\partial t}\right)_a, \quad (18)$$

where

$$H_s = \hbar 2\omega_0 A_{ee} + \hbar(\omega_0 + \Omega_{12})A_{ss} + \hbar(\omega_0 - \Omega_{12})A_{aa} \tag{19}$$

is the Hamiltonian of the interacting atoms,

$$\left(\frac{\partial\rho}{\partial t}\right)_s = -\frac{\Gamma + \Gamma_{12}}{2}\{(A_{ee} + A_{ss})\rho + \rho(A_{ee} + A_{ss}) - 2(A_{se} + A_{gs})\rho(A_{es} + A_{sg})\} \tag{20}$$

describes dissipation through the symmetric transition $|e\rangle \to |s\rangle \to |g\rangle$, and

$$\left(\frac{\partial\rho}{\partial t}\right)_a = -\frac{\Gamma - \Gamma_{12}}{2}\{(A_{ee} + A_{aa})\rho + \rho(A_{ee} + A_{aa}) - 2(A_{ae} + A_{ga})\rho(A_{ea} + A_{ag})\} \tag{21}$$

describes dissipation through the antisymmetric transition $|e\rangle \to |a\rangle \to |g\rangle$. The symmetric transition decays with the rate $\Gamma + \Gamma_{12}$, while the antisymmetric transition decays with the rate $\Gamma - \Gamma_{12}$. For $-\pi/2 < \omega_0\tau < \pi/2$, the decay rate of the symmetric (antisymmetric) transition increases (decreases), whereas for $\pi/2 < \omega_0\tau < 3\pi/2$ the decay rate of the symmetric (antisymmetric) transition decreases (increases). Furthermore, for the extreme value $\kappa = 1$, which corresponds to an ideal optical element, the symmetric and antisymmetric states are completely decoupled from the environment when $\omega_0\tau = (2n+1)\pi$ and $\omega_0\tau = 2n\pi$, respectively, and therefore they can be regarded as decoherence-free states. Under these conditions, the two-atom system under consideration reduces to a three-level cascade system that is similar to the small-sample model or two-atom Dicke model [53] for a system of two close atoms without optical element [42]. The major difference is that in a system of two close atoms only the decoupling of the antisymmetric state from the environment is possible [42], whereas in the system of two distant atoms coupled via an optical element, depending on the interatomic distance, the occurrence of decoupling from the environment is possible not only for the antisymmetric state, but also for the symmetric state.

Now we assume that initially the density matrix of the system of two distant atoms has the so-called 'X' form [54,55]

$$\rho = \begin{pmatrix} \rho_{11} & 0 & 0 & \rho_{14} \\ 0 & \rho_{22} & \rho_{23} & 0 \\ 0 & \rho_{32} & \rho_{33} & 0 \\ \rho_{41} & 0 & 0 & \rho_{44} \end{pmatrix}. \tag{22}$$

Physically, the 'X' form corresponds to a situation where all coherences between the ground state $|1\rangle \equiv |g_1, g_2\rangle$ and the single excitation states $|2\rangle \equiv |g_1, e_2\rangle$ and $|3\rangle \equiv |e_1, g_2\rangle$, and between $|2\rangle, |3\rangle$ and the double excitation state $|4\rangle \equiv |e_1, e_2\rangle$ are zero. It should be noted that the density matrix with 'X' form in the bare states basis, has also 'X' form in the collective basis and this form is preserved during the evolution governed by the master equation. The equations of motion for the non-zero elements of the density matrix in the collective basis are as follows

$$\begin{aligned}\frac{\partial\rho_{ee}}{\partial t} &= -2\Gamma\rho_{ee}, \frac{\partial\rho_{eg}}{\partial t} = -\Gamma\rho_{eg}, \\ \frac{\partial\rho_{ss}}{\partial t} &= -(\Gamma + \Gamma_{12})(\rho_{ss} - \rho_{ee}), \\ \frac{\partial\rho_{aa}}{\partial t} &= -(\Gamma - \Gamma_{12})(\rho_{aa} - \rho_{ee}), \\ \frac{\partial\rho_{as}}{\partial t} &= -(\Gamma + 2i\Omega_{12})\rho_{as}. \end{aligned} \tag{23}$$

By solving Equations (23), we find all the matrix elements required for calculating the time evolution of the system, in particular, the entanglement evolution. The solutions of the above equations can be easily obtained as follows

$$\begin{cases} \rho_{ee}(t) = \rho_{ee}(0)e^{-2\Gamma t}, \\ \rho_{eg}(t) = \rho_{eg}(0)e^{-\Gamma t}, \end{cases} \text{for all } t \tag{24}$$

$$\begin{cases} \rho_{ss}(t) = \rho_{ss}(0)e^{-\Gamma t} + \rho_{ee}(0)e^{-\Gamma t}(1 - e^{-\Gamma t}) & t < \tau, \\ \rho_{ss}(t) = \rho_{ss}(0)e^{-\Gamma(t+\kappa\cos(\omega_0\tau)(t-\tau))} \\ \quad + \rho_{ee}(0)e^{-\Gamma t}\left\{\frac{(1+\kappa\cos(\omega_0\tau))}{(1-\kappa\cos(\omega_0\tau))}(e^{-\Gamma(\tau+\kappa\cos(\omega_0\tau)(t-\tau))} - e^{-\Gamma t}) + (1 - e^{-\Gamma\tau})e^{-\Gamma\kappa\cos(\omega_0\tau)(t-\tau)}\right\} & t > \tau, \end{cases} \tag{25}$$

$$\begin{cases} \rho_{aa}(t) = \rho_{aa}(0)e^{-\Gamma t} + \rho_{ee}(0)e^{-\Gamma t}(1 - e^{-\Gamma t}) & t < \tau, \\ \rho_{aa}(t) = \rho_{aa}(0)e^{-\Gamma(t-\kappa\cos(\omega_0\tau)(t-\tau))} + \rho_{ee}(0)e^{-\Gamma t}\left\{\frac{(1-\kappa\cos(\omega_0\tau))}{(1+\kappa\cos(\omega_0\tau))}(e^{-\Gamma(\tau-\kappa\cos(\omega_0\tau)(t-\tau))} - e^{-\Gamma t}) + (1 - e^{-\Gamma\tau})e^{\Gamma\kappa\cos(\omega_0\tau)(t-\tau)}\right\} & t > \tau, \end{cases} \tag{26}$$

$$\begin{cases} \rho_{as}(t) = \rho_{as}(0)e^{-\Gamma t} & t < \tau, \\ \rho_{as}(t) = \rho_{as}(0)e^{-\Gamma(t+i\kappa\sin(\omega_0\tau)(t-\tau))} & t > \tau. \end{cases} \tag{27}$$

### 3.2. Entanglement dynamics

We are now in a position to investigate the dynamical behavior of the atom–atom entanglement in the system under consideration. Several different measures have been proposed to identify entanglement between two atoms, and we choose the Wootters entanglement measure [56], the concurrence $C$, defined as

$$C(t) = \max\{0, \sqrt{\lambda_1} - \sqrt{\lambda_2} - \sqrt{\lambda_3} - \sqrt{\lambda_4}\}, \quad (28)$$

where $\lambda_1, ..., \lambda_4$ are the eigenvalues in decreasing order of magnitude of the 'spin-flipped' density operator $R(t) = \rho(t)\tilde{\rho}(t)$. The matrix $\rho(t)$ is the density matrix for the two atoms and the matrix $\tilde{\rho}(t)$ is defined by

$$\tilde{\rho}(t) = (\sigma_y \otimes \sigma_y)\rho^*(t)(\sigma_y \otimes \sigma_y), \quad (29)$$

where $\sigma_y$ is the Pauli matrix and $\rho^*(t)$ is the complex conjugation of $\rho(t)$. The range of the concurrence is from 0 to 1. For unentangled atoms $C(t) = 0$, whereas $C(t) = 1$ for the maximally entangled atoms. In the case of the density matrix with the 'X' form, the concurrence could be easily calculated by [22]

$$C(t) = \max\{0, C_1(t), C_2(t)\}, \quad (30)$$

where

$$C_1(t) = 2(|\rho_{14}| - \sqrt{\rho_{22}\rho_{33}}), \quad (31)$$

and

$$C_2(t) = 2(|\rho_{23}| - \sqrt{\rho_{11}\rho_{44}}). \quad (32)$$

In the collective basis, $C_1(t)$ and $C_2(t)$ take the following forms

$$C_1(t) = 2|\rho_{eg}(t)| \\ - \sqrt{(\rho_{ss}(t) + \rho_{aa}(t))^2 - (2\mathrm{Re}\rho_{as}(t))^2}, \quad (33a)$$

$$C_2(t) = \sqrt{(\rho_{ss}(t) - \rho_{aa}(t))^2 + (2\mathrm{Im}\rho_{as}(t))^2} \\ - 2 \times \sqrt{\rho_{ee}(t)\rho_{gg}(t)}. \quad (33b)$$

In the following, we shall concentrate on two specific cases: (1) when initially only one of the atoms is excited, and (2) when initially both atoms are excited.

#### 3.2.1. Initial state with one atom excited

In this case, the initial state is separable, so there is no atom–atom entanglement initially (initial one-photon unentangled state). In terms of the collective states we have $\rho_{ee}(0) = 0$ and $\rho_{ss}(0) = \rho_{aa}(0) = \rho_{as}(0) = 1/2$. By using Equations (24)–(27), the nonzero elements of the density matrix read

$$\rho_{ss}(t) = \rho_{aa}(t) = \rho_{as}(t) = \frac{1}{2}e^{-\Gamma t} \quad (t < \tau), \quad (34)$$

$$\begin{cases} \rho_{ss}(t) = \frac{1}{2}e^{-\Gamma(t + \kappa \cos(\omega_0 \tau)(t - \tau))}, \\ \rho_{as}(t) = \frac{1}{2}e^{-\Gamma(t + i\kappa \sin(\omega_0 \tau)(t - \tau))}, \\ \rho_{aa}(t) = \frac{1}{2}e^{-\Gamma(t - \kappa \cos(\omega_0 \tau)(t - \tau))}. \end{cases} (t > \tau) \quad (35)$$

As the above equations show, the density matrix is of the 'X' form in which the populations of the excited and the ground states as well as two photon coherences are zero. Hence, the concurrence can be calculated by Equation (30). It is clear that $C_1$ is negative and for positive values of $C_2$ the concurrence is equal to $C_2$, i.e. $C(t) = \max\{0, C_2(t)\}$, where

$$\begin{cases} C_2(t) = 0 \quad t \leq \tau, \\ C_2(t) = e^{-\Gamma t}\sqrt{\sinh^2(\kappa\Gamma\cos(\omega_0\tau)(t - \tau)) + \sin^2(\kappa\Gamma\sin(\omega_0\tau)(t - \tau))} \quad t > \tau. \end{cases} \quad (36)$$

For $t < \tau$, the two atoms are disentangled but after the time $\tau$ the appearance of collective effects may lead to ESB. The birth time of entanglement can be controlled by changing the interatomic distance. Another interesting feature of the entanglement in the system under consideration is the dependence of the concurrence on the properties of the optical element ($\kappa$). In Figure 2 we have plotted the concurrence for atom–atom entanglement as a function of the dimensionless evolution time $\Gamma t$ for $\omega_0 \tau = (2n + 1)\pi/2$ ($n$ integer), i.e. $\Gamma_{12} = 0, \Omega_{12} \neq 0$, and for different values of the parameter $\kappa$. As is seen, there is no entanglement at early times of evolution, but suddenly after the time $\tau$ an entanglement emerges (ESB). Furthermore, we realize that the entanglement increases with increasing the parameter $\kappa$. After evolving for a finite time, the atom-atom entanglement decreases such that for small and intermediate values of $\kappa$ it decays exponentially in time (no sudden death), while for large values of $\kappa$, ESD appears, i.e. the concurrence decays in non-exponential way and vanishes at a finite time. As can be seen, ESD is followed by a revival of entanglement, after which the asymptotic decay of entanglement takes place.

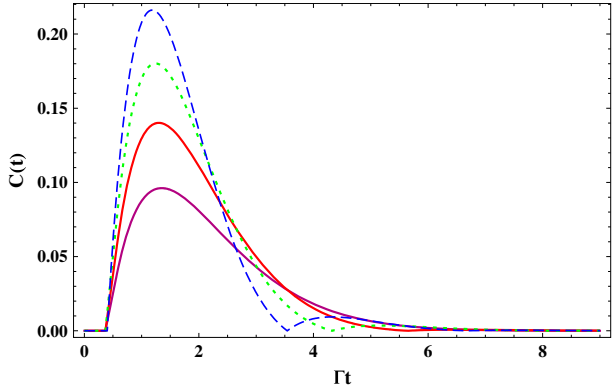

Figure 2. Time evolution of concurrence $C(t)$ as a function of $\Gamma t$ for $\omega_0 \tau = \frac{(2n+1)\pi}{2}$, $\Gamma \tau = 0.4$, and for different values of the parameter $\kappa$: $\kappa = 0.4$ (solid line), $\kappa = 0.6$ (dashed-dotted line), $\kappa = 0.8$ (dotted line), and $\kappa = 1$ (dashed line). (The color version of this figure is included in the online version of the journal.)

Figure 3 illustrates the behavior of the concurrence as a function of $\Gamma t$ for $\omega_0 \tau = n\pi$ ($n$ integer), and for different values of $\kappa$. In this case, there is no dipole–dipole interaction between the two atoms, $\Omega_{12} = 0$, but the collective damping is nonzero, $\Gamma_{12} \neq 0$. It can be seen that in this case, after the time $\tau$, ESB takes place and the entanglement increases with increasing the parameter $\kappa$. However, there is no ESD; for $\kappa < 1$ the atom–atom entanglement that has already been created decays exponentially in time, while for $\kappa = 1$ it remains in the system and attains its maximal value, $C \approx 0.35$, as time goes on. For $\omega_0 \tau = n\pi$, the two-atom system behaves as an effective three-level atom in which the symmetric and the antisymmetric states decay with the rates $\Gamma(1 + (-1)^n \kappa)$ and $\Gamma(1 - (-1)^n \kappa)$, respectively. As Equation (33b) shows, the concurrence depends on the population difference $\rho_{ss}(t) - \rho_{aa}(t)$. For even values of $n$, $\rho_{ss}(t)$ decays faster than $\rho_{aa}(t)$, and eventually only the antisymmetric state survives. For long times, as Figure 4 shows, the concurrence which decays with a reduced rate $\Gamma(1 - \kappa)$, is equal to the population $\rho_{aa}(t)$. For odd values of $n$, the situation is reversed; $\rho_{aa}(t)$ decays faster than $\rho_{ss}(t)$ and for the case of an ideal optical element ($\kappa = 1$) the symmetric state is totally decoupled from the vacuum modes. Thus, population becomes trapped in the symmetric state. As mentioned before, this is a major difference between the system of two distant atoms coupled via an optical element and a system of two close atoms, in which only the decoupling of the antisymmetric state from the environment is possible.

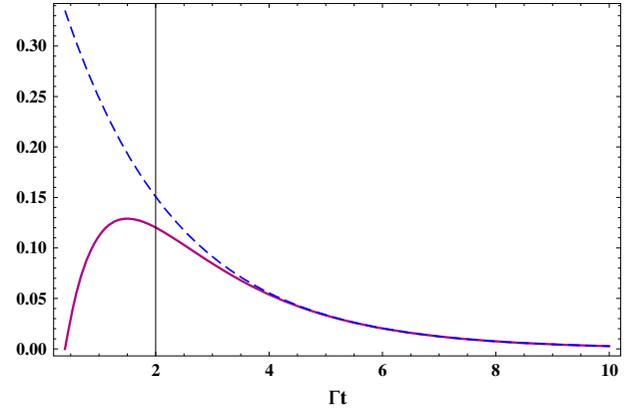

Figure 4. Time evolution of concurrence $C(t)$ (solid line) and the population of the antisymmetric state $\rho_{aa}(t)$ (dashed line) for $\kappa = 0.5$, $\Gamma \tau = 0.4$, and $\omega_0 \tau = 2n\pi$. (The color version of this figure is included in the online version of the journal.)

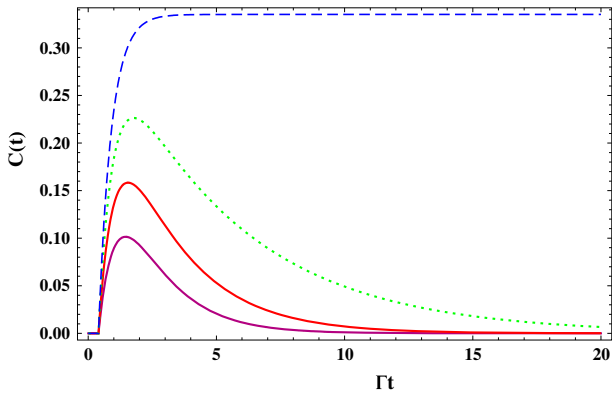

Figure 3. Time evolution of concurrence $C(t)$ as a function of $\Gamma t$ for $\omega_0 \tau = n\pi$, $\Gamma \tau = 0.4$, and for different values of the parameter $\kappa$: $\kappa = 0.4$ (solid line), $\kappa = 0.6$ (dashed-dotted line), $\kappa = 0.8$ (dotted line), and $\kappa = 1$ (dashed line). (The color version of this figure is included in the online version of the journal.)

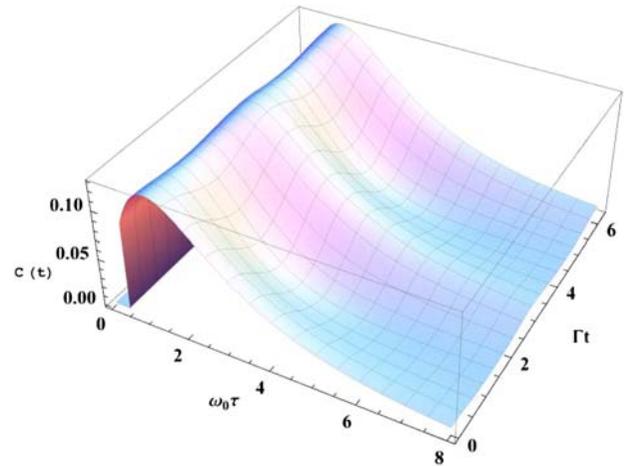

Figure 5. The concurrence $C(t)$ as a function of $\Gamma t$ and $\omega_0 \tau$ for $\kappa = 0.5$ and $\Gamma \tau = 0.4$ when initially only one of the atoms is excited. (The color version of this figure is included in the online version of the journal.)

Figure 5 displays the concurrence versus the dimensionless evolution time $\Gamma t$ and the interatomic distance. As can be seen, suddenly after the time $\tau$ a transient entanglement is created between the two atoms. The amount of atom-atom entanglement and the time interval in which it remains in the system depend strongly on the distance between the atoms.

### 3.2.2. Initial state with two atoms excited

Another interesting example of the atom–atom entanglement evolution takes place when both atoms are prepared initially in their excited states (two-photon unentangled state), so the only non-zero element of the atomic density matrix is $\rho_{ee}(0) = 1$. By using Equations (24)–(27), the nonzero elements of the density matrix are obtained as

$$\begin{cases} \rho_{ee}(t) = e^{-2\Gamma t}, \\ \rho_{ss}(t) = \rho_{aa}(t) = e^{-\Gamma t}(1 - e^{-\Gamma t}), \end{cases} t < \tau \quad (37)$$

$$\begin{cases} \rho_{ee}(t) = e^{-2\Gamma t}, \\ \rho_{ss}(t) = e^{-\Gamma t}\{\frac{(1+\kappa\cos(\omega_0\tau))}{(1-\kappa\cos(\omega_0\tau))}(e^{-\Gamma(\tau+\kappa\cos(\omega_0\tau)(t-\tau))} - e^{-\Gamma t}) + (1 - e^{-\Gamma\tau})e^{-\Gamma\kappa\cos(\omega_0\tau)(t-\tau)}\}, t > \tau \\ \rho_{aa}(t) = e^{-\Gamma t}\{\frac{(1-\kappa\cos(\omega_0\tau))}{(1+\kappa\cos(\omega_0\tau))}(e^{-\Gamma(\tau-\kappa\cos(\omega_0\tau)(t-\tau))} - e^{-\Gamma t}) + (1 - e^{-\Gamma\tau})e^{\Gamma\kappa\cos(\omega_0\tau)(t-\tau)}\}. \end{cases} \quad (38)$$

Since in this case $\rho_{eg}(t) = 0$, according to Equation (33a), $C_1(t)$ is negative and the only contribution to the atom–atom entanglement, if any, can come from $C_2(t)$, which is given by

$$\begin{cases} C_2(t) = 0 \quad t \leq \tau, \\ C_2(t) = e^{-\Gamma t}\left[e^{-\Gamma\tau}(\frac{(1+\kappa\cos(\omega_0\tau))}{(1-\kappa\cos(\omega_0\tau))}e^{-\Gamma\kappa\cos(\omega_0\tau)(t-\tau)} - \frac{(1-\kappa\cos(\omega_0\tau))}{(1+\kappa\cos(\omega_0\tau))}e^{\Gamma\kappa\cos(\omega_0\tau)(t-\tau)})\right. \\ \left. + 2(1 - e^{-\Gamma\tau})\sinh(\Gamma\kappa\cos(\omega_0\tau)(t-\tau)) - \frac{4\kappa\cos(\omega_0\tau)}{(1-\kappa^2\cos^2(\omega_0\tau))}e^{-\Gamma t}\right] t > \tau. \end{cases} \quad (39)$$

It is clear from Equation (33b) that entanglement can result solely from unequal populations of the symmetric and antisymmetric states. In Figure 6 we have plotted the time evolutions of the concurrence, $C(t) = \max\{0, C_2(t)\}$, together with the populations $\rho_{ee}(t)$ and $\rho_{aa}(t)$ of the excited and the antisymmetric states, for initially both atoms excited, with $\omega_0\tau = 2n\pi$. As can be seen, there is no entanglement at early times of evolution, but suddenly after the excited state becomes depopulated, the concurrence builds up (delayed ESB) and then it decays asymptotically to zero with the decay rate of the antisymmetric state. The reason for the delayed creation of this transient entanglement can be understood by noting that during the evolution, the system decays first from the initial excited state to the symmetric and antisymmetric states. Since for $\omega_0\tau = 2n\pi$ the decay rate of the symmetric state, $\Gamma(1 + \kappa)$, is faster than that of the antisymmetric state, $\Gamma(1 - \kappa)$, there appears unbalanced population distribution between these states resulting in an entanglement between the two atoms. At time when the symmetric state becomes depopulated, the state of the system reduces to the maximally entangled antisymmetric state. This explains why at later times the evolution of the concurrence follows the evolution of the population of the antisymmetric state. It should be noted for $\omega_0\tau = (2n + 1)\pi$ the atom–atom entanglement, similar to the case of $\omega_0\tau = 2n\pi$, occurs when the excited state becomes depopulated with the only difference that the concurrence decays asymptotically to zero with the decay rate of the symmetric state since the antisymmetric state decays faster than the symmetric state.

Figure 7 displays the concurrence versus the dimensionless evolution time $\Gamma t$ and the interatomic distance when both atoms are prepared initially in their excited states. As can be seen, the amount of the entanglement as well as its decay rate depend on the interatomic distance such that only for those distances that the collective damping rate $\Gamma_{12}$ is different from zero the transient atom-atom entanglement is created in the system. By comparison with the result obtained for the case in which the atomic system is initially prepared in a one-photon unentangled state, we find that the delayed ESB from the initially two-photon unentangled state has different origin.

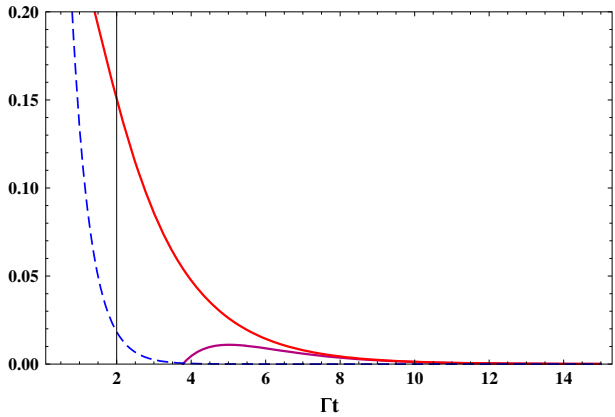

Figure 6. Time evolution of $C(t)$ (solid line), $\rho_{ee}(t)$ (dashed line), and $\rho_{aa}(t)$ (dashed-dotted line) for initially both atoms excited with $\kappa = 0.5$, $\Gamma\tau = 0.4$, and $\omega_0\tau = 2n\pi$. (The color version of this figure is included in the online version of the journal.)

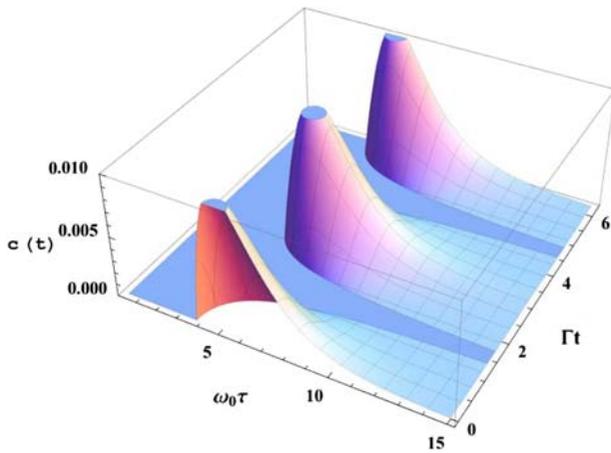

Figure 7. The concurrence $C(t)$ as a function of $\Gamma t$ and $\omega_0\tau$ for $\kappa = 0.5$ and $\Gamma\tau = 0.4$ when initially both atoms are excited. (The color version of this figure is included in the online version of the journal.)

In the former case, ESB may appear only after the time $\tau$, whereas in the latter case, it appears when the excited state becomes depopulated. The depopulation time of the excited state is longer than the time $\tau$ and it is uncontrollable, whereas the time $\tau$ can be controlled through the change of the distance between the two atoms.

## 4. Summary and conclusions

In summary, we have studied the sudden features of atom–atom entanglement for a system of two distant two-level atoms coupled via an optical element and interacting with a common environment being in the vacuum state. We have shown that the presence of an optical element between two distant atoms leads to the revival of collective effects. Particularly, in contrast to the system of two close atoms without optical element, where only the decoupling of the antisymmetric state from the environment is possible, in the system under consideration depending on the interatomic distance, both symmetric and antisymmetric states can become decoherence-free states. We have studied the dynamics of the atom–atom entanglement for two specific initial conditions. We have found that the evolution of the entanglement is sensitive not only to the interatomic distance but also to the initial state of the system as well as to the properties of the optical element. For the initial condition with only one of the atoms is excited, if $\omega_0\tau = (2n+1)\pi/2$, i.e. when $\Gamma_{12} = 0$ and $\Omega_{12} \neq 0$, the entanglement can be created abruptly after the finite time $\tau$ (delayed ESB). After evolving for a finite time, the entanglement decreases such that for small and intermediate values of $\kappa$ there is no ESD, while for large values of $\kappa$, ESD appears that is followed by a revival of entanglement, after which the asymptotic evolution of entanglement takes place. On the other hand, for $\omega_0\tau = n\pi$, i.e. when $\Gamma_{12} \neq 0, \Omega_{12} = 0$, after the time $\tau$, delayed ESB takes place but there is no ESD; for $\kappa < 1$ the atom–atom entanglement decays exponentially in time, while for $\kappa = 1$ it remains in the system and attains its maximal value. For the case in which both atoms are prepared initially in their excited states we have found that delayed ESB takes place when the excited state becomes depopulated. For $\omega_0\tau = 2n\pi$, at later times the evolution of the entanglement follows the asymptotic decay of the population of the antisymmetric state while for $\omega_0\tau = (2n+1)\pi$ it decays asymptotically to zero with the decay rate of the symmetric state.

Finally, we want to point out that the formalism used in this paper can be extended to more than two atoms. The simplest multi-atom case is a three-atom system. In [57], an Ising-type atom–atom interaction in the cavity QED system has been obtained by connecting three distant cavities via optical fibers. The authors have discussed the generation of remote two-atom and three-atom entanglements governed by this interaction. They have found that the overall two-atom (bipartite) entanglement is very small because of the existence of the third atom. However, the three-atom (tripartite) entanglement has a much longer period than two-atom entanglement and can reach a maximum very close to 1. Therefore, it is expected that our treatment may be applied to investigate the abovementioned features of bipartite and tripartite entanglements, particularly related to ESB and ESD, in a system of three distant atoms without cavity, coupled via an appropriate lens combination. We hope to report on such an issue in a forthcoming paper.

## Acknowledgements

The authors would like to express their sincere thanks to the referee for useful suggestions. They are also grateful to the Office of Graduate Studies of the University of Isfahan for their support.

**Appendix. Derivation of the reduced master Equation (14)**

To derive the reduced master Equation (14) we first change the sums over $\mu$ and $\nu$ in Equation (12) into integrals,

$$\sum_\mu g_\mu^2 e^{i(\omega_0-\omega_\mu)(t-t')} \rightarrow \frac{1}{(2\pi)^3 \varepsilon_0 \hbar c^3} \int_0^\infty d\omega \omega^3 e^{i(\omega_0-\omega)(t-t')}$$
$$\times \int_{4\pi-\delta\Omega_0} d\Omega(p^2 - |\vec{p}\cdot\vec{k}|^2/k^2), \qquad (40)$$

$$\sum_{\nu} g_{\nu}^2 e^{i\vec{k}_{\nu}.(\vec{r}_i-\vec{r}_j)} e^{i(\omega_0-\omega_{\nu})(t-t')} \rightarrow \frac{1}{2(2\pi)^3 \varepsilon_0 \hbar c^3}$$

$$\times \int_0^{\infty} d\omega \omega^3 e^{i(\omega_0-\omega)(t-t')}$$

$$\times \int_{\delta\Omega_0} d\Omega e^{i\vec{k}.(\vec{r}_i-\vec{r}_j)} (p^2 - |\vec{p}.\vec{k}|^2/k^2)$$

$$= \frac{\kappa |\vec{p}|^2}{6\pi^2 \varepsilon_0 \hbar c^3} \times \int_0^{\infty} d\omega \omega^3 e^{i(\omega_0-\omega)(t-t')} e^{i\omega\tau} \ (i \neq j), \quad (41)$$

where we have made the replacement $e^{i\vec{k}.(\vec{r}_i-\vec{r}_j)} \rightarrow e^{i\omega\tau}$ with $\tau = d/c$ being the time a photon emitted inside the solid angle $\delta\Omega_0$ needs to cover the distance between one atom and the other via the optical element. We now carry out the integral over $d\Omega$ in Equation (40). For this purpose we assume that the dipole moments of the atoms are parallel and oriented in the $z$-direction. In the spherical coordinates $\vec{k} = k(\sin\theta\cos\varphi, \sin\theta\sin\varphi, \cos\theta)$ where $\theta$ is the angle between $\vec{r}_i - \vec{r}_j$ and $\vec{k}$. In this representation, the unit polarization vectors can be chosen as

$$\hat{e}_{\vec{k}_1} = (-\cos\theta\cos\varphi, -\cos\theta\sin\varphi, \sin\theta),$$
$$\hat{e}_{\vec{k}_2} = (\sin\varphi, -\cos\varphi, 0). \quad (42)$$

Thus, we have

$$\sum_{\mu} g_{\mu}^2 e^{i(\omega_0-\omega_{\mu})(t-t')} \rightarrow \frac{|\vec{p}|^2}{6\pi^2 \varepsilon_0 \hbar c^3} \int_0^{\infty} d\omega \omega^3 e^{i(\omega_0-\omega)(t-t')}. \quad (43)$$

Under the Markov approximation, we can evaluate the integral over $t'$, appearing in Equation (12) to obtain

$$\lim_{t\to\infty} \int_0^t dt' \rho(t-t') e^{ixt'} \approx \rho(t)[\pi\delta(x) + iP(1/x)], \quad (44)$$

where P indicates the principal value of the integral. By using Equations (41), (43), and (44), the reduced master Equation (12) takes the following form

$$\frac{\partial \rho}{\partial t} = -\sum_{i,j=1,2} \frac{\Gamma_{ij}}{2}(\rho(t)\sigma_i^+(t)\sigma_j^-(t) + \sigma_i^+(t)\sigma_j^-(t)\rho(t)$$
$$-2\sigma_i^-(t)\rho(t)\sigma_j^+(t)) - i\sum_{i\neq j}^{2} \Omega_{ij}[\sigma_i^+(t)\sigma_j^-(t), \rho(t)], \quad (45)$$

where

$$\Gamma_{ii} \equiv \Gamma = \frac{|\vec{p}|^2 \omega_0^3}{3\hbar\pi\varepsilon_0 c^3},$$
$$\Gamma_{ij} = \kappa\Gamma\cos(\omega_0\tau)\Theta(t-\tau) \ (i\neq j),$$
$$\Omega_{ij} = \frac{\kappa\Gamma}{2}\sin(\omega_0\tau)\Theta(t-\tau) \ (i\neq j). \quad (46)$$